\documentstyle[sprocl,epsfig]{article}
\epsfclipon

\def\be{\begin{equation}}
\def\ee{\end{equation}}
\def\bea{\begin{eqnarray}}
\def\eea{\end{eqnarray}}

\begin{document}

\title{METRIC PERTURBATIONS AND INFLATIONARY PHASE TRANSITIONS}

\author{D. CORMIER}

\address{Institute for Physics, University of Dortmund,\\ 
D-44221 Dortmund, Germany\\E-mail: cormier@dilbert.physik.uni-dortmund.de} 

\author{R. HOLMAN}

\address{Physics Department, Carnegie Mellon University,\\
Pittsburgh PA\ 15213, U.S.A.\\E-mail: holman@fermi.phys.cmu.edu}


\maketitle\abstracts{We study the out of equilibrium dynamics of 
inflationary phase transitions and compute the resulting spectrum
of metric perturbations relevant to observation.  We show that 
simple single field models of inflation may produce an adiabatic 
perturbation spectrum with a blue spectral tilt and that the precise
spectrum depends on initial conditions at the outset of inflation.}

\section{Field Evolution}
We work in a spatially flat Friedmann-Robertson-Walker universe
with scale factor $a(t)$ and take the inflaton to be a real scalar field 
with Lagrangian
\begin{equation}
L = \frac12 \nabla^\mu \Phi \nabla_\mu \Phi -
\left[\frac{3 \mu^4}{2 \lambda} - \frac{1}{2} \mu^2 \Phi^2 + \frac{\lambda}{4!}
\Phi^4\right] \; .
\end{equation}

We break up the field $\Phi$ into its expectation
value, defined within the closed time path formalism,~\cite{ctp} and 
fluctuations about that value:
$$
\Phi(\vec{x},t) = \phi(t) + \psi(\vec{x},t) \; , \; \;
\phi(t)  \equiv  \langle \Phi(\vec{x},t) \rangle \; .
$$

By imposing a Hartree resummation,~\cite{hartree} we arrive at the following
equations of motion for the inflaton:~\cite{thesis}
\begin{equation}
\ddot{\phi} + 3\frac{\dot{a}}{a}\dot{\phi} - \mu^2 \phi 
+ \frac{\lambda}{6}\phi^3 + \frac{\lambda}{2}\langle\psi^2\rangle \phi  
=  0 \; ,
\label{phieqn} 
\end{equation}
\begin{equation}
\left[\frac{d^2}{dt^2} + 3\frac{\dot{a}}{a}\frac{d}{dt} + 
\frac{k^2}{a^2} - \mu^2 + \frac{\lambda}{2}\langle\psi^2\rangle 
+ \frac{\lambda}{2}\phi^2  \right] f_k = 0 \; .
\label{fkeqn}
\end{equation}
The fluctuation $\langle \psi^2 \rangle$ is determined 
from the mode functions $f_k$:
\begin{equation}
\langle \psi^2 \rangle = \int \frac{d^3k}{(2\pi)^3} |f_k|^2 
\; .
\label{fluctuation}
\end{equation}
For $a(t_0)=1$, the initial conditions on the mode functions are
\begin{equation}
f_k(t_0) = \frac{1}{\sqrt{2\omega_k}} \; , \; \;
\dot{f}_k(t_0) = \left(-\dot{a}(t_0) - 
i\omega_k \right) f_k(t_0) \; ,
\label{fkin}
\end{equation}
with
$$
\omega_k^2 \equiv k^2 - \mu^2 + \frac{\lambda}{2}\langle\psi^2\rangle 
+ \frac{\lambda}{2}\phi^2 - \frac{R(t_0)}{6} \; .
$$
$R(t_0)$ is the initial Ricci scalar.
For small $k$, we modify $\omega_k$ either by
means of a quench or by explicit deformation so that the frequecies
are real.~\cite{grav,desitter}

The gravitational dynamics are determined by the semi-classical
Einstein equation.~\cite{birrell} For a minimally coupled inflaton we have
\begin{eqnarray}
\frac{\dot{a}^2}{a^2} &=& \frac{8\pi G_N}{3} 
\left[\frac12 \dot{\phi}^2 + \frac12 \langle \dot{\psi}^2 \rangle
+ \frac{1}{2a^2} \left\langle (\vec{\nabla}\psi)^2 \right\rangle 
\right. \nonumber \\
&+& \left. \frac{3 \mu^4}{2\lambda} - \frac12 \mu^2 
\left(\phi^2+\langle\psi^2\rangle\right)
+ \frac{\lambda}{24}\left(\phi^4+3\langle\psi^2\rangle^2
+6\phi^2 \langle\psi^2\rangle\right) \right] \; ,
\label{frweqn}
\end{eqnarray}
where $G_N$ is Newton's gravitational constant, and 
\begin{eqnarray}
\langle \dot{\psi}^2(t) \rangle &\equiv& \int \frac{d^3k}{(2\pi)^3}
|\dot{f}_k|^2  \; , 
\label{psidot} \\
\left\langle \left(\vec{\nabla} \psi(t) \right)^2 \right\rangle &\equiv&
\int \frac{d^3k}{(2\pi)^3} k^2 |f_k|^2 \; .
\label{nablapsi}
\end{eqnarray}
Each of these integrals is regulated using a cutoff with the divergences 
absorbed into a renormalization of the parameters of the theory.~\cite{grav}

\begin{figure}[t]
\epsfig{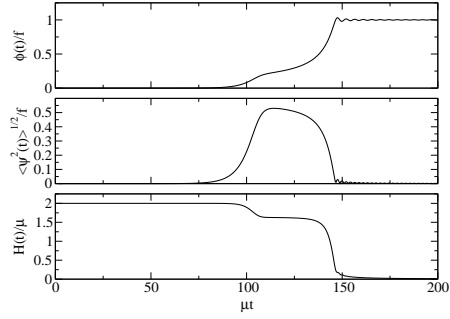}
\caption{The mean field $\phi(t)/f$, the fluctuation $<\psi^2(t)>^{1/2}/f$, 
  and the Hubble parameter $H(t)/\mu$ vs. $t$ 
  with $\phi(t_0) = 0.4 H_0/2\pi$, $\dot{\phi}(t_0)=0$, $H_0=2\mu$,
  $\lambda/8\pi^2=10^{-16}$, and $f \equiv \mu\sqrt{6/\lambda}$.}
\label{psh2}
\end{figure}

A typical field evolution is depicted in Fig.~\ref{psh2}.

\section{Metric Perturbations}
Following the procedure of Mukhanov, Feldman and Brandenberger,~\cite{mfb} 
we arrive at the expression for the density contrast at mode 
re-entry:\cite{spinodal}
\begin{eqnarray}
|\delta_h(k)| & = & \sqrt{\frac{2}{75\pi}} \frac{1}{M_{Pl} 
\left(\dot{\phi}^2 + 
\langle \dot{\psi}^2 \rangle\right)} \left[ \frac{3 \mu^4}{2\lambda} 
- \frac12 \mu^2 \left(\phi^2+\langle\psi^2\rangle\right)\right. \nonumber \\
& & \left. + \frac{\lambda}{24}\left(\phi^4+6\phi^2\langle\psi^2\rangle 
+3\langle\psi^2\rangle^2\right)\right]^{1/2} \nonumber \\
& \times & \left[\mu^4 \left(\phi^2+\langle\psi^2\rangle\right) 
- \frac{\lambda \mu^2}{3}\left(\phi^4 + 6\phi^2\langle\psi^2\rangle 
+ 3 \langle\psi^2\rangle^2
\right)\right. \nonumber \\
& & \left. + \frac{\lambda^2}{36} \left(\phi^6 + 15\phi^4\langle\psi^2\rangle
+ 45\phi^2\langle\psi^2\rangle^2 + 15\langle\psi^2\rangle^3\right)\right]^{1/2}
\; .
\label{lphi4deltaH}
\end{eqnarray}
The computation of the tilt parameter $n_s-1$ is straightforward,
given (\ref{lphi4deltaH}):  
\begin{equation}
n_s-1 \equiv \left. \frac{d (\ln|\delta_h(k)|)}{d \ln(k)}\right|_{k=aH} \; .
\label{nsminusone}
\end{equation}

As gravitational wave perturbations do not directly interact with the 
inflaton field, they may be related directly to the expansion rate.  
The amplitude of gravitational waves is simply:~\cite{mfb}
\begin{equation}
|\delta_g(k)| = \frac{2}{\sqrt{3\pi}} \frac{H}{M_{Pl}} \; ,
\label{deltaG}
\end{equation}
All expressions are to be evaluated when the given scale $k$ first crosses 
the horizon, $k = a H$.

\begin{figure}[t]
\epsfig{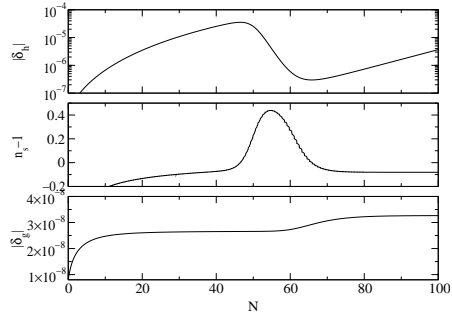}
\caption{The scalar amplitude $\delta_h$, the scalar tilt $n_s-1$,
  and the tensor amplitude $\delta_g$ as a function of the number of 
  $e$-folds $N$ before the end of inflation that the scale first crosses 
  the horizon. Parameters are as in Fig.~\ref{psh2}.}
\label{dnd2}
\end{figure}

\begin{figure}[t]
\epsfig{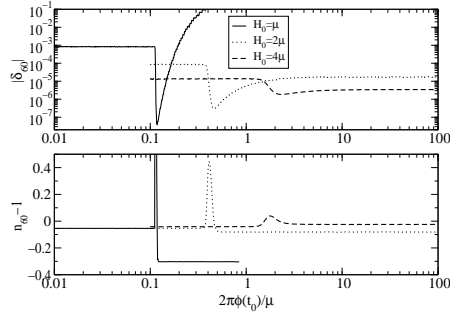}
\caption{The scalar amplitude $\delta_{60}$ and tilt 
  $n_{60}-1$ of the scale crossing the horizon $60$ $e$-folds
  before the inflation ends vs. $2\pi\phi(t_0)/\mu$
  with $\dot{\phi}(t_0)=0$, $\lambda/8\pi^2=10^{-16}$ and several 
  values of $H_0$.}
\label{deltaphi}
\end{figure}

An example perturbation spectrum is shown in Fig.~\ref{dnd2}, while 
Fig.~\ref{deltaphi} shows the dependence of the spectrum on the initial
state for a number of possible evolutions.  Both of these figures show
distinct regions characterized by a blue spectral tilt.

\section*{Acknowledgments}
R.H. was supported in part by the Department of Energy Contract 
DE-FG02-91-ER40682.


\end{document}